\documentclass[twocolumn,showpacs,preprintnumbers,prC]{revtex4}
\usepackage{graphicx}
\def\bea {\begin{eqnarray}}
\def\eea {\end{eqnarray}}

\def\be {\begin{equation}}
\def\ee {\end{equation}}

\begin{document}
\title{Effects of equation of state on nuclear suppression and 
the initial entropy density of quark gluon plasma}
\author{Surasree Mazumder and Jan-e Alam}
\medskip
\affiliation{Theoretical Physics Division, 
Variable Energy Cyclotron Centre, 1/AF, Bidhan Nagar ,
Kolkata - 700064}
\date{\today}
\begin{abstract}
We study the effects of the equation of state on the
nuclear suppression of heavy flavours in quark gluon plasma and
estimate the initial entropy density of the system  
produced at the highest RHIC energy.  
For this purpose we have used the experimental 
data on the charged particle multiplicity and the nuclear suppression of 
single electron spectra originating from the semi-leptonic decays of open 
charm and beauty mesons. We have used inputs from 
lattice QCD to  minimize the model dependence of the results. 
We obtain the value of the 
initial entropy density which varies from 20 to 59 /fm$^3$ depending on the 
value of the velocity of sound that one uses for the analysis. Our 
investigation leads to a conservative value of the initial entropy density 
$\sim 20/$fm$^3$ with corresponding initial temperature $\sim 210$ MeV 
well above the value of the transition temperature predicted by lattice QCD.   
\end{abstract}

\maketitle
A thermalized system of quarks and gluons, called quark gluon plasma (QGP)
is expected to be formed in
the collisions of two nuclei at ultra-relativistic energies~\cite{npa2005}. 
Rigorous 
experimental and theoretical efforts are on to create and characterize
this novel, deconfined phase of quarks and gluons.
Lattice QCD (LQCD) calculations indicate that at a temperature $\sim 175$ MeV
the entropy density ($s$) of the hadronic matter 
rises significantly due to the release 
of colour degrees of freedom which are confined within the hadrons at zero temperature.
Therefore, it is of foremost importance to determine the value of the initial 
entropy density ($s_i$) / initial temperature ($T_i$) 
for the system formed in nuclear collisions at Relativistic
Heavy Ion Collider (RHIC) and Large Hadron Collider (LHC) and assess 
whether the system is formed in colour deconfined phase or not. 
The focus of the present study 
is to estimate $s_i$ or $T_i$ of the 
system formed at Au+Au collisions at $\sqrt{s_{\mathrm NN}}=200$ GeV.
For this purpose we strict to take inputs from  experimental
data and LQCD calculations to minimize the model dependence  
of the outcome of the present analysis. 

One of the
possible way to estimate the value of the initial entropy density is the 
extrapolation of the  measured (final) observables backward in time
through a suitable dynamical model. 
In absence of  viscous loss the time reversal symmetry of the system is valid,
therefore, the measured multiplicity at the freeze-out of the system
can be used to estimate $s_i$.
The $s_i$  and the thermalization time
($\tau_i$) are constrained by the  measured (final) hadron multiplicity
($dN/dy$)  by the following relation~\cite{hwa}:
\be
s_i\tau_i=\kappa\frac{1}{A_\perp}\frac{dN}{dy}
\label{dndy}
\ee
where $A_\perp$ is the transverse area of the system can be
determined from the collision geometry and
$\kappa$ is a known constant (=3.7 for massless bosons like pions).
The value of $dN/dy$ which is connected to
$s_i$ through Eq.~\ref{dndy} is readily available 
for different collision centralities~\cite{npa2005}.
In Eq.~\ref{dndy} there are two unknown
quantities, $\tau_i$ and $s_i$ both of which can not be determined
from a single equation involving  a single measured the $dN/dy$. Therefore,
we choose another experimentally measured quantity the nuclear suppression of 
heavy quarks (HQ), 
$R_{\mathrm AA}$~\cite{stare,phenixe}, which is sensitive to the 
initial condition  
and hence is very useful to estimate $s_i$.

The advantage of choosing the heavy flavour are (a) they are produced 
in early hard collisions and hence can witness the
entire evolution of the QGP and (b) HQ are 
Boltzmann suppressed at the temperature range 
expected to be achieved in heavy ion collisions at RHIC, 
therefore, the HQ do not determine the bulk features of the QGP.
The magnitude of $R_{\mathrm AA}$  depends on the
amount of drag the heavy quark faces during its propagation
through QGP.  The heavy quarks being the witness of the 
early condition and  the
drag (and diffusion) coefficient is
a temperature dependent quantity that makes the $R_{\mathrm AA}$  
a good probe for the  measurement of initial temperature.

Earlier various attempts have been made to
explain the experimental results on $R_{AA}$ -  some of
these are addition of non-perturbative
effects from the quasi-hadronic bound state~\cite{hvh},
three-body scattering process~\cite{ko},
dissociation of open heavy flavoured mesons  by the
thermal partons ~\cite{adil}  and inclusion of
temperature dependent strong coupling~\cite{gossiaux}.

We briefly outline the procedure 
of evaluating the $R_{AA}$ for  single electrons
originating from the semi-leptonic decays of heavy mesons
produced from the fragmentation of the HQ. 
The HQ while propagating through the QGP dissipates 
energy in the medium and hence its momentum gets 
attenuated. The magnitude of the momentum degradation
gets reflected in the experimentally measured quantity, $R_{AA}$
mentioned above. 
Theoretically the momentum evolution of the HQ in the
expanding QGP background can be described by using
Fokker Planck Equation (FPE)~\cite{sc,rapp,turbide,bjoraker,svetitsky,japrl,
npa1997,munshi,rma,surasree,santosh}. 
The evolution of the probe {\it i.e.} the HQ is described by the FPE:
\be
\frac{\partial f}{\partial t}= \frac{\partial}{\partial p_{i}}\left[A_{i}(\textbf{p})f+\frac{\partial}{\partial p_{j}}[B_{ij}
(\textbf{p})f]\right]~~,
\label{landaukeq}
\ee
where
\be
A_{i}= \int d^{3}\textbf{k}w(\textbf{p},\textbf{k})k_{i}= \gamma p_i~~,
\label{eqdrag}
\ee
and
\be
B_{ij}= \frac{1}{2} \int d^{3}\textbf{k}w(\textbf{p},\textbf{k})k_{i}k_{j}=D\delta_{ij}.
\label{eqdiff}
\ee
$\gamma$ and D are called the drag and diffusion coefficients,
contain the interaction of the probe with the medium. 
To solve the FPE we need to supply the drag and diffusion coefficients
and the initial HQ (charm and beauty) momentum distributions.

There are two main processes through which the HQ dissipates 
energy in the QGP: 
(i) energy dissipation can take place due to
the elastic collisions of the HQ with the 
quarks, antiquarks and gluons in the thermal bath,
(ii) the radiative process due to which 
the HQ  emits soft gluons (which subsequently  
get absorbed in the QGP)  due to its interaction
with the QGP. The details solution of FPE with
temperature and momentum 
dependent drag and diffusion coefficients including both the processes 
(i) and (ii) have been discussed in our earlier works~\cite{surasree}.
The initial momentum distributions
of charm and bottom quarks at RHIC energy ($\sqrt{s_{\mathrm NN}}=200$ GeV)
have been taken from MNR code~\cite{MNR}.
The method of solving the FPE numerically with temperature and 
momentum dependent transport coefficients including other issues
are discussed in ~\cite{surasree}, therefore we do not repeat those
details here.

Classically the induced radiation takes place due to the
jiggling motion of the propagating particle in the medium.
Since the heavier particle jiggle less consequently 
induced energy loss
is expected to be smaller [dead cone effect~\cite{DK} 
(see also~\cite{wang,adsw})] 
for HQ compared to light particles.
However, the experimental data from RHIC indicates similar
amount of energy loss by heavy quarks and light partons 
in the measured kinematic range.  Various reasons  like
the anomalous mass dependence of the radiative process 
due to the finite size of the QGP ~\cite{zakharov}, 
development of dead cone due to high virtuality of
the partons resulting from the dismantling of colour
fields during the initial hard collisions~\cite{bzk}     
have been proposed as reasons for this observation.
The authors in~\cite{roy} concluded that 
the reduction in the energy loss of HQ due to radiative
process is
due to the dead cone effect but it is fair to mention
that the issue is yet to be settled.

The other effect which influences the radiative loss
is the Landau-Pomeranchuk-Migdal(LPM) effect. 
This effect originate due to the interplay between
two time  scales of the system~\cite{klein}: the formation time  
($\tau_F$) and the mean scattering time ($\tau_{sc}$) 
of the gluon radiated from the HQ. LPM effect imposes
certain constrain on the phase space of the emitted
gluon~\cite{wgp,dks}. Both the  dead cone and the LPM effects have been
taken in to account in evaluating
the drag and diffusion coefficients 
in the present work (see~\cite{surasree,santosh} 
for details).
  
\begin{figure}[h]
\begin{center}
\includegraphics[scale=0.43]{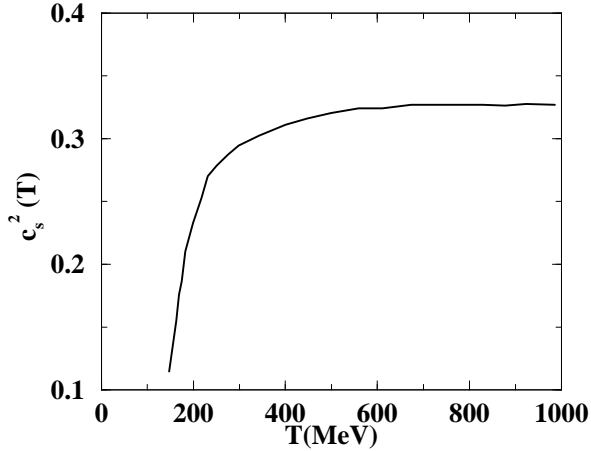}
\caption{Velocity of sound squared as a function temperature~\cite{lattice}}
\label{fig1}
\end{center}
\end{figure}

Now we discuss the QGP background with which
the HQ interacts. 
The equation of state (EoS) plays a crucial role 
in describing the space time evolution of the expanding
QGP from the initial state to the quark-hadron 
transition point. We use boost invariant hydrodynamic model~\cite{bjorken}
with the  LQCD calculation EoS ~\cite{lattice} for the space time
description of the matter.
The velocity of sound ($c_s$) as obtained in ~\cite{lattice} 
from LQCD calculations shows a significant variation with temperature
(Fig.~\ref{fig1}).
It starts with a very low value of $c_s^2$ at $T\sim T_c$ and then increases
with $T$ to reach the maximum value ($c_s^2=1/3$) corresponding to the
value of a massless ideal gas. The EoS for almost baryon free QGP expected 
at RHIC energy is take as: $P=c_s^2\epsilon$. The EoS sets the expansion time
scale for the system as $\tau_{\mathrm exp}\sim [(1/\epsilon)d\epsilon/d\tau]
^{-1}\sim\tau/(1+c_s^2)$  indicating the fact that lower value of $c_s$ 
makes the expansion time scale longer {\it i.e.} the rate of expansion slower.
Therefore,  for given values of $T_i$ and $T_c$ the life time of the
QGP will be longer for smaller $c_s$. The value of $T_c$ is fixed 
at 175 MeV.

We solve the FPE numerically~\cite{surasree}  for momentum 
dependent drag and diffusion coefficients 
to get the charm and beauty quarks momentum distribution. 
The solution then convoluted with the fragmentation functions~\cite{peterson}
to obtain the transverse momentum distribution of the $D$ and $B$ mesons 
which subsequently decay to create leptons~\cite{gronau,ali}.

In the same way the lepton spectra from the heavy flavours produced in 
p-p collisions can be calculated from the charm and beauty quark 
distributions which enter as initial conditions to the FPE.
The solution of FPE contains the effects of drag (quenching) 
on the HQ where as the the initial distributions of HQ 
does not contain any such effects, therefore the ratio 
of these two quantities, the nuclear suppression factor,
$R_{AA}$ act as a marker for the medium. This 
is observed experimentally through the depletion of 
$R_{AA}$ at high transverse momentum ($p_T$).  

\begin{figure}
\begin{center}
\includegraphics[scale=0.43]{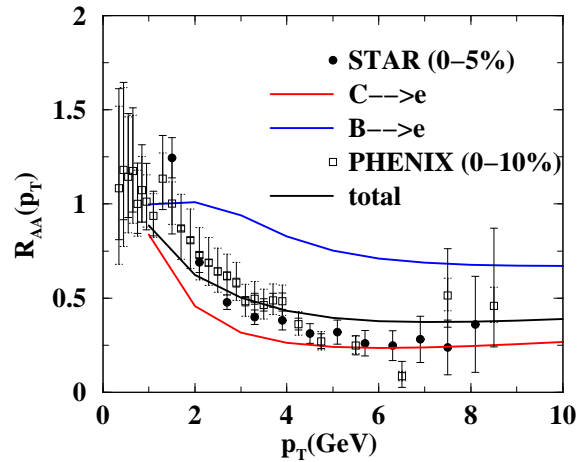}
\caption{(colour online) Variation of $R_{AA}$  with
$p_T$ for $c_s^2=1/3$ and  $T_i=300$ MeV.}
\label{fig2}
\end{center}
\end{figure}

As discussed above for larger $c_s$ the 
expansion time scale is shorter {\it i.e.} 
the QGP life time is smaller. consequently 
the HQ spends less time in the QGP  which
ultimately leads to less suppression of
the single electron spectra originating 
from the decays of HQ. Therefore, we take
the following strategy to obtain the allowed
range of initial temperature. We take the highest
possible value of $c_s^2 (=1/3)$ for the space-time
description of the flowing QGP background, in the present 
approach this will lead to the maximum value of  $T_i$. 
In this case the HQ will spend the lesser
amount of time in the QGP. Therefore, to 
achieve the experimentally measured $R_{\mathrm AA}$ 
one will need larger drag or in other word larger
initial temperature. The results for $c_s^2=1/3$
is displayed in Fig.~\ref{fig2}. The value of $T_i$
obtained from the analysis for this case is $300$ MeV,
the corresponding value of $s_i=2\pi^2g_{\mathrm eff}T_i^3/45\sim 59/$
fm$^{3}$. The value of $g_{\mathrm eff}\sim 38$ is extracted
from the variation of 
$s/T^3$ with $T$ provided by the LQCD calculations~\cite{lattice}.

\begin{figure}
\begin{center}
\includegraphics[scale=0.43]{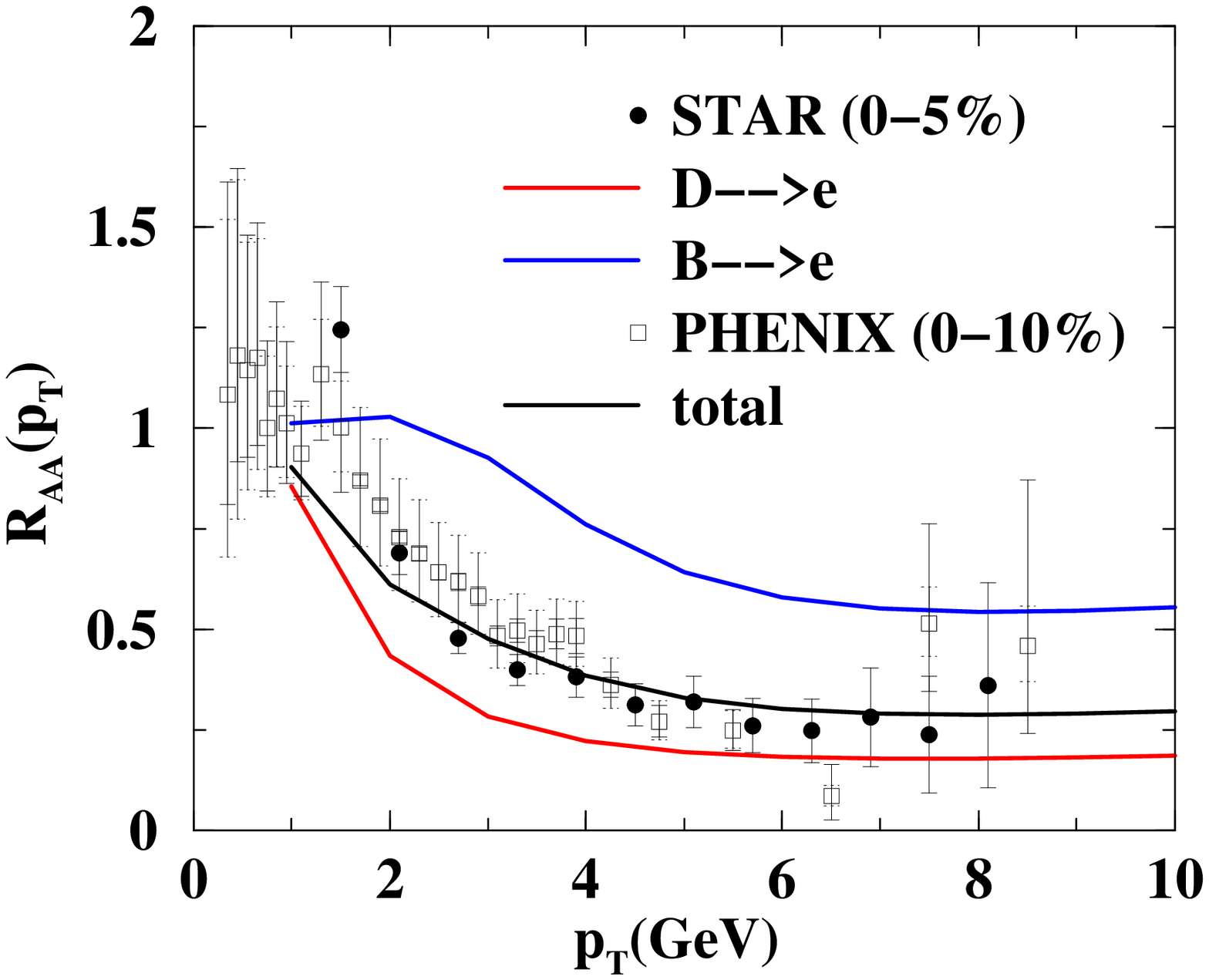}
\caption{(colour online) Variation of $R_{AA}$ with $p_T$ for 
for $c_s^2=1/4$ and $T_i=240$ MeV.}
\label{fig3}
\end{center}
\end{figure}

In Fig.~\ref{fig3} results for $c_s^2=1/4$ 
are  depicted. For $c_s^2=1/4$ the HQ spend longer time
in QGP than for $c_s^2=1/3$.  Therefore, with a lower initial
temperature, $T_i=240$ MeV and $s_i\sim 29.34/$fm$^{3}$, the 
data can be reproduced.

For $c_s^2=1/5$
the HQ spend longer time in QGP
(compared to the case when $c_s^2=1/4$). 
Therefore, the observed suppression 
dictates to reduce the initial
temperature. In this case the data is well 
reproduce with $T_i=210$ MeV and $s_i\sim 19.66/$fm$^3$.

Further lowering of $c_s$ will make the value of $\tau_i$ large 
(for given $dN/dy$) enough to contradict other results like
the observation of large hadronic elliptic flow which requires
small $\tau_i$ (see~\cite{v2flow} for review).  That will also results in lower $T_i$ with
which it will be difficult explain other experimental results.

\begin{figure}
\begin{center}
\includegraphics[scale=0.43]{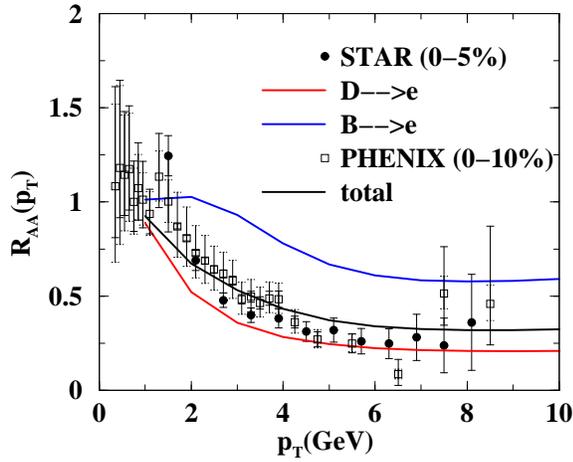}
\caption{(colour online) Variation of $R_{AA}$ with $p_T$ for $c_s^2=1/5$ 
and $T_i=210$ MeV.}
\label{fig4}
\end{center}
\end{figure}
\begin{figure}[h]
\begin{center}
\includegraphics[scale=0.43]{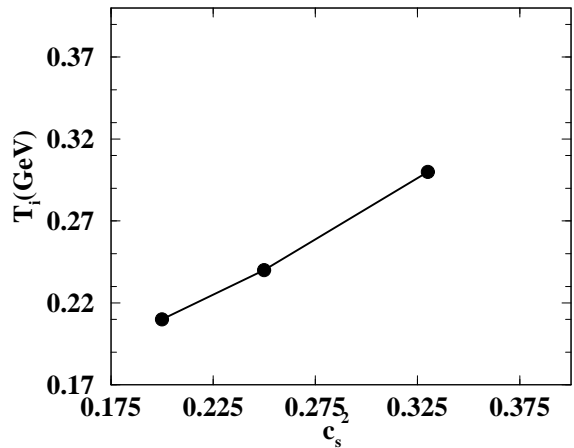}
\caption{The variation of $T_i$ with $c_s^2$ for fixed $dN/dy$.}
\label{fig5}
\end{center}
\end{figure}

For all the theoretical results displayed in Figs.~\ref{fig2},~\ref{fig3} and
~\ref{fig4} we have kept the quantity $dN/dy$ constant consequently
the value of $\tau_i$ changes to 0.6, 1.17 and 1.7 fm/c 
for $T_i=300, 240$ and $210$ MeV
respectively. The changes in $T_i$ is forced by the change in the EoS.
In Fig.~\ref{fig5} we show the variation of $T_i$ with $c_s^2$
obtained by constraints imposed by the 
experimental data on $R_{\mathrm AA}$ and $dN/dy$.
The value of $T_i$ varies from 210 to 300 MeV. 
In this context we compare the value of $T_i$ 
obtained in the work  with some of those  reported
earlier.  In Refs.~\cite{hvh} the value of $T_i$ 
is obtained as $\sim 375$  MeV from the study of
heavy quark suppression. From the simultaneous analysis
of light and heavy quarks suppressions in Ref.~\cite{prl105}
a value of $T_i=400$ MeV is obtained.  
The authors in Ref.~\cite{jpg32,prc84,jpg34} mentioned the values of
the initial gluon rapidity distribution, $dN_g/dy$, 
which may be converted to $T_i=290, 270$ and $310$ respectively.

It is interesting 
to note that the lowest value of $T_i$ obtained from the present
analysis is well above the quark-hadron phase transition temperature, 
indicating the fact that the system  formed in Au+Au collisions
at $\sqrt{s_{\mathrm NN}}=200$ GeV might be formed 
in the partonic phase. 

In summary, we have studied the effects of the EoS on the
suppression of single electrons originating 
from the decays of heavy flavours produced in Au+Au collisions
at $\sqrt{s_{\mathrm NN}}=200$ GeV. We found that the initial temperature
may vary from 210 to 300 MeV depending on the velocity of sound, which
sets the scale for the expansion that one uses. We have used experimental data
(charged particle multiplicity and
$R_{\mathrm AA}$ of heavy flavours) and LQCD results 
($c_s$, $g_{\mathrm eff}$ etc.) to keep the model dependence minimum.  
The effects of transverse expansion is neglected here. 
With the transverse expansion the HQ will (1) travel longer path
(2) with diluted density.  However, the two 
competing effects (1) and (2) will have some sort of cancellation
due to which our final conclusion may not get altered.
 
\noindent{Acknowledgement:} SM is grateful to Department of Atomic Energy,
Government of India for financial support. We are thankful to Santosh K Das
and Trambak Bhattacharyya for useful discussions.


\begin{thebibliography}{100}

\bibitem{npa2005} I. Arsene {\it et al.} (BRAHMS Collaboration),
Nucl. Phys. A {\bf 757}, 1 (2005); B. B. Back {\it et al.} (PHOBOS
Collaboration), Nucl. Phys. A {\bf 757}, 28 (2005);
J. Adams {\it et al.} (STAR Collaboration), Nucl. Phys. A {\bf 757}, 102 (2005);
K. Adcox {\it et al.} (PHENIX Collaboration),
Nucl. Phys. A {\bf 757}, 184,(2005).

\bibitem{hwa} R. C. Hwa and K. Kajantie, Phys. Rev. D, {\bf 32}, 1109 (1985). 

\bibitem{stare} B. I. Abeleb {\it et al.} (STAR Collaboration), Phys. Rev.
Lett. {\bf 98}, 192301 (2007).

\bibitem{phenixe} S. S. Adler {\it et al.} (PHENIX Collaboration),
Phys. Rev. Lett. {\bf 96}, 032301 (2006).

\bibitem{hvh}  H. van Hees, M. Mannarelli, V. Greco and R. Rapp,
Phys. Rev. Lett. {\bf 100}, 192301 (2008).

\bibitem{ko} C. M. Ko and W. Liu, Nucl. Phys. A {\bf 783}, 23c (2007).

\bibitem{adil} A. Adil and I. Vitev, Phys. Lett. B  {\bf 649}, 139 (2007).

\bibitem{gossiaux}  P. B. Gossiaux and J. Aichelin, Phys. Rev. C
{\bf 78}, 014904 (2008).

\bibitem{sc} S. Chakraborty and D. Syam, Lett. Nuovo Cim. {\bf 41}, 381 (1984).

\bibitem{rapp} H. van Hees, R. Rapp, Phys. Rev. C,{\bf 71}, 034907 (2005).


\bibitem{turbide} S. Turbide, C. Gale, S. Jeon and G. D. Moore,
Phys. Rev. C {\bf 72}, 014906 (2005).

\bibitem{bjoraker} J. Bjoraker and R. Venugopalan, Phys. Rev. C {\bf 63},
024609 (2001).

1\bibitem{svetitsky} B. Svetitsky, Phys. Rev. D {\bf 37}, 2484( 1988).

\bibitem{japrl} J. Alam, S. Raha and B. Sinha, Phys. Rev. Lett. {\bf 73}, 1895
(1994).

\bibitem{npa1997} P. Roy, J. Alam, S. Sarkar, B. Sinha and S. Raha,
Nucl. Phys. A {\bf 624}, 687 (1997).

\bibitem{munshi} M  G. Mustafa and  M. H. Thoma, Acta Phys. Hung. A
{\bf 22}, 93 (2005).

\bibitem{rma} P. Roy, A. K. Dutt-Mazumder and J. Alam, Phys. Rev. C {\bf 73},
044911 (2006).

\bibitem{surasree} S. Mazumder, T. Bhattacharya, J. Alam and
S. K. Das, arXiv:1106.2615 [nucl-th]; Phys. Rev. C (in press).

\bibitem{santosh} S. K. Das, J. Alam and P. Mohanty,
Phys. Rev. C {\bf 82}, 014908 (2010);
S. K. Das, J. Alam and P. Mohanty, Phys. Rev. C {\bf 80},  054916 (2009);
S. K. Das, J. Alam, P. Mohanty and B. Sinha,
Phys. Rev. C {\bf 81}, 044912 (2010).

\bibitem{MNR} M. L. Mangano, P. Nason and G. Ridolfi, Nucl. Phys.B {\bf 373}, 295 (1992).

\bibitem{DK} Y. L. Dokshitzer and D. E. Kharzeev, Phys. Lett. B, {\bf 519},
199 (2001).

\bibitem{wang} B. W. Zhang, E. Wang and X.-N. Wang, Phys. Rev. Lett.
{\bf 93}, 072301 (2004).

\bibitem{adsw} N. Armesto, A. Dainese, C. A. Salgado and U. A. Wiedemann,
Phys. Rev. D {\bf 71}, 054027 (2005).


\bibitem{zakharov} B. G Zakharov, JETP Lett. {\bf 86}, 444(2007);
P. Aurenche and B. G. Zakharov, JETP Lett. {\bf 90}, 237 (2009).

\bibitem{bzk} B. Z. Kopeliovich, I. K. Potashnikova, I. Schmidt,
Phys. Rev. C  {\bf 82}, 037901(2010)

\bibitem{roy} Roy A. Lacey {\it et al.}, Phys. Rev. Lett.{\bf 103},142302(2009)


\bibitem{klein} S. Klein, Rev. Mod. Phys. {\bf 71}, 1501 (1999)

\bibitem{wgp} X. N. Wang, M. Gyulassy and M. Pl\"umer,
Phys. Rev. D {\bf 51}, 3436 (1995).


\bibitem{dks}
M. G. Mustafa, D. Pal, D. K. Srivastava and M. H. Thoma, Phys. Lett. B, {\bf 428},234(1998).


\bibitem{bjorken}J. D. Bjorken, Phys. Rev. D {\bf 27}, 140 (1983).

\bibitem{lattice} S. Bors\'ayi {\it et al.}, 
J. High. Ener. Phys {\bf 1011}, 077 (2010)

\bibitem{peterson} C. Peterson {\it et al.}, Phys. Rev. D {\bf 27}, 105 (1983).

\bibitem{gronau} M. Gronau, C. H. Llewellyn Smith,
T. F. Walsh, S. Wolfram and T. C. Yang,
Nucl. Phys. B {\bf 123}, 47 (1977).

\bibitem{ali} A. Ali, Z. Phys. C {\bf 1}, 25 (1979).


\bibitem{v2flow} P. Huovinen and P. V. Ruuskanen,
Ann. Rev. Nucl. Part. Sci. {\bf 56} (2006) 163;
D. A. Teaney, arXiv:0905.2433 [nucl-th].


\bibitem{prl105} G. Qin and A. Majumder, Phys. Rev. Lett. {\bf 105}, 262301 (2010).

\bibitem{jpg32} M. Djordjevic, J. Phys. G {\bf 32}, 333 (2006).

\bibitem{prc84} J. Uphoff, O. Fochler, Z. Xu and C. Greiner, Phys. Rev. C {\bf 84}, 024908(2011).

  
\bibitem{jpg34} I. Vitev, A. Adil and H. Van Hees, J. Phys. G {\bf 34}, 769 (2007).

\end{thebibliography}
\end{document}